\documentclass{elsart}

\usepackage{natbib}
\usepackage{graphicx}
\usepackage{amssymb}
\usepackage{amsmath}
\usepackage{bm}

\newcommand{\g}{{\rm \, g}}
\newcommand{\s}{{\rm \, s}}
\newcommand{\cm}{{\rm \, cm}}
\newcommand{\mm}{{\rm \, mm}}
\newcommand{\pc}{{\rm \, pc}}
\newcommand{\ipc}{{\rm \, pc^{-3}}}
\newcommand{\km}{{\rm \, km}}
\newcommand{\kms}{{\rm \, km~s^{-1}}}

\newcommand{\gcm}{{\rm \, g~cm^{-2}}}
\newcommand{\yr}{{\rm \, yr}}
\newcommand{\Gyr}{{\rm \, Gyr}}
\newcommand{\Myr}{{\rm \, Myr}}
\newcommand{\au}{{\rm \, AU}}
\newcommand{\ergs}{{\rm \, ergs~s^{-1}}}
\newcommand{\K}{{\rm \, K}}
\newcommand{\GHz}{{\rm \, GHz}}
\newcommand{\THz}{{\rm \, THz}}

\newcommand{\apj}{ApJ}
\newcommand{\aap}{A\&A}

\newcommand{\mnras}{MNRAS}
\newcommand{\aj}{AJ}
\newcommand{\prd}{PRD}
\newcommand{\bain}{Bull. Astro. Inst. Netherlands}
\newcommand{\baas}{BAAS}

\newcommand{\nat}{{\it Nature}}

\newcommand{\araa}{ARA\&A}

\begin{document}

\begin{frontmatter}



\title{Imprint of Distortions in the Oort Cloud on the CMB Anisotropies}
\author{Daniel Babich\thanksref{label1}}
\ead{babich@tapir.caltech.edu}
\author{Abraham Loeb\thanksref{label2}}

\address[label1]{ California Institute of Technology, Theoretical Astrophysics, MC 130-33
Pasadena, CA 91125, USA}
\address[label2]{Astronomy Department, Harvard University, 60 Garden Street, Cambridge, MA
02138, USA}

\begin{abstract}
We study the effect of a close encounter of a passing star on the shape of the inner Oort Cloud, using the 
impulse approximation. The deviation of the perturbed Oort Cloud from sphericity adds angular fluctuations 
to the brightness of the Cosmic Microwave Background (CMB) due to thermal emission by the comets. An encounter 
with a solar-mass star at an impact parameter of $2100 \au$, as expected based on the abundance and velocity 
dispersion of stars in the solar neighborhood, leads to a quadrupole moment $C_2 = 4.5 (3.5) \times 10^{-15}, 
6.7 (1.1) \times 10^{-12}, 1.1 (0.11) \times 10^{-9}$ at $\nu = 30, 353, 545 \GHz$, respectively in intensity 
and (temperature) fluctuations. We also quantify the quadrupole spectral distortions produced by the Scattered 
Disc, which will exist irregardless of any perturbation and the subsequent shape of the Oort Cloud. For 
comparison, the temperature fluctuation quadrupole moment predicted by the current cosmological model is 
$C_2 = 1.76 \times 10^{-10}$, which corresponds to fluctuation in the CMB intensity of $C_2 = 2.9 \times 10^{-10}, 
6.8 \times 10^{-9}, 1.6 \times 10^{-8}$ at $\nu = 30, 353, 545 \GHz$.  Finally, we discuss how a measurement of 
the anisotropic spectral distortions could be used to constrain the trajectory of the closest stellar fly-by.
\end{abstract}

\begin{keyword}
cosmic microwave background \sep Oort Cloud
96.50.Hp \sep 98.70.Vc
\end{keyword}

\end{frontmatter}

\section{Introduction}
\label{sec:intro}

At the outer edge of the Solar System exists a swarm of small icy rocks known as the Oort Cloud \citep{Oort50,Dones04}. 
Comets with long orbital periods, $P \geq 200 \yr$, are ejected from this cloud into the inner Solar System by the 
perturbations from passing stars, giant molecular clouds or tides from the Galactic disc. The perturbed comets are moved 
onto nearly-parabolic orbits which approach the Sun, develop comae and may be detected. Gravitational perturbations by 
the planets can strongly alter the comets' orbits as they enter the inner Solar System, and so typically they either get 
ejected from the Solar System or left strongly bound to the Sun. It is very unlikely that a comet will return again on a 
nearly parabolic orbit. The observed continual flux of long period comets on nearly parabolic orbits originally led \citet{Oort50} 
to propose that all long period comets were entering the inner Solar System for the first time and therefore a resevoir of 
comets must exists in the outermost region of the Solar System.

There is considerable interest in studying the Oort Cloud since it is a remnant of the formation epoch of the Solar System. 
This is true for both the individual objects, whose chemical composition may reveal information about the composition and 
thermal state of the outer regions of the proto-planetary disc (but see \citet{Mumma93} for a comprehensive review of
processes that may cause significant evolution in these properties over the past $4 \Gyr$) as well as the dynamical structure 
of the Oort Cloud as a whole. The cloud structure, which is largely determined by its formation scenario, may retain information 
about the masses of the giant planets, the rate at which passing stars perturbed the comets' orbits and the surface density of 
any gas that was potentially still within the Solar System. This information could provide important clues about the formation 
processes of the outer planets.

The Oort Cloud is cruedly divided into an inner and an outer region. This division stems from the fact that the probability for 
ejection into the inner Solar System by external perturbations is a strong function of semi-major axis \citep{Heisler86,Dones04}. 
All of the observed long-period comets are believed to have originated from the outer Oort Cloud. The perturbations that eject 
comets into the inner Solar System also isotropize the comets' orbits and eliminates any information about the formation epoch
in the structure of the outer Oort Cloud. Therefore the reduced sensitivity of the comets in the inner Oort Cloud to these external 
perturbers is both good and bad. On the one hand it leads us to suspect that the structure of the inner Oort Cloud should retain 
an imprint of its formation as well as any rare, strong gravitational scattering event that might have occured over the past 4.5 
billion years. On the other hand, it makes studying the inner Oort Cloud difficult. It is impossible to directly observe the comets
in the inner Oort Cloud through their reflected sunlight because they are too far. Extremely large objects like Sedna can be optically 
detected when they are close to perihelion \citep{Brown04}, however it will be difficult to detect the potentially large number of small 
objects ($R \le 1 \km$) in this region. One novel technique involves using the microwave emission from these bodies to constrain, or 
possibly detect, the bulk properties of objects in this region \citep{Babich06}.

At a distance $D$ from the sun an Oort Cloud object will be heated by the Sun to an equilibrium temperature,
\begin{align}
T_{\rm Oort} = & \left( \frac{(1-A) L_{\odot}}{16 \epsilon \sigma \pi D^2}\right)^{1/4} \\ 
  \simeq & 8.5 \K \sqrt{\frac{1000 \au}{D}},
\end{align}
here we adopted an albedo of $A = 4 \%$ \citep{Jewitt98} and an emissivity of $\epsilon \simeq 1.0$. We adopt a model such that the emissivity
of an Oort Cloud object is approximately unity when its size is larger than the wavelength of the relevant radiation and zero when the
wavelength is larger. Also, $\sigma$ is the Stefan-Boltzmann constant and $L_{\odot} = 3.83 \times 10^{33} \ergs$ is the Solar luminosity. 
The equilibrium temperature is higher than the Cosmic Microwave Background (CMB) temperature of $T_{\rm CMB} = 2.725 \K$ \citep{Mather99}. 
When the comets are farther away 
from the Sun a variety of physical processes help maintain temperatures greater than $5-6 \K$ \citep{Mumma93}. The Oort Cloud objects 
extinguish some fraction of the CMB radiation and emit blackbody radiation at a higher temperature \citep{Babich06}. The observed intensity 
of the CMB radiation is then
\begin{equation}
I_{\nu}(\hat{\bm n}) = [1-\tau(\hat{\bm n})]B_{\nu}[T_{\rm CMB}(\hat{\bm n})] 
+ \tau(\hat{\bm n}) B_{\nu}[T_{\rm Oort}(\hat{\bm n})],
\end{equation}
where $\tau$ is the optical depth and $T_{\rm Oort}$ is temperature of the Oort Cloud objects along the direction $\hat{\bm n}$. 
\citet{Babich06} constrained the mean, or monopole, contribution to the CMB spectrum using the {\it Cosmic Background Explorer Far Infrared 
Absolute Spectrophotometer} (COBE FIRAS) data \citep{Fixsen97,Mather99} and placed upper limits on the total mass in the inner Oort Cloud for a variety 
of models of the size distribution of Oort Cloud objects. In this paper we will focus on effects that can make the Oort Cloud non-spherical 
and subsequently produce anisotropic spectral distortions in the CMB. 

We should emphasize that the induced spectral distortions considered in this paper are not the chemical potential ($\mu$) or Compton-y spectral 
distortion that are commonly discussed in the literature. The signal produced in the outer Solar System is that of a sum of weighted blackbodies 
at different temperatures, which in general does not have a blackbody frequency spectrum. The exception is the Rayleigh-Jean's portion of the 
spectrum, where the weighted sum of blackbodies does produce a frequency spectrum described by a Rayleigh-Jean's distribution at 
a weighted temperature. This implies that low frequency data will not be very useful in allowing us to detect a signal originating from the 
outer Solar System. See \citet{Babich06} for a discussion of techniques to distinguish the spectral distortions produced in the outer Solar 
System from those caused by the CMB temperature anisotropies. 

The Oort Cloud is believed to be nearly spherical based on the statistics of the inclination angles of new long period 
comets \citep{Marsden71}. The observed apsides of these comets are not uniformly distributed across the sky, however the 
observed clustering trends are believed to be due to the structure of the Galactic potential \citep{Delsemme87,Dones04}. 
The Oort Cloud has been populated by objects that formed in the Sun's proto-planetary disc and were scattered multiple 
times by the four giant planets into increasingly higher eccentricity and higher energy orbits. This process continued 
until some external perturbation lifted the comets' perihelia well beyond Neptune's orbit, and then the comets became
decoupled from the inner Solar System. The radial distribution of the Oort Cloud comets, which is set by how far the gas 
giant planets can scatter the comets before their perihelia is lifted safely outside the inner Solar System, strongly depends 
on the external environment of the young Solar System. The evolutionary processes that lifted the perihelia of comets
beyond the planetary region of the Solar System must have also made shperical the distribution of comets which where 
initially in the Ecliptic Plane.  
Simulations of this formation process have found an inner edge of the Oort Cloud, defined as 
the semi-major axis at which the comets are isotropically distributed, near $a = 3000 \au$ \citep{Duncan87} and $a = 5000 \au$ 
\citep{Dones07}. It should be emphasized that objects with smaller semi-major axes exist, but may be more confined towards the
Ecliptic Plane. This is the Scattered Disc which smoothly extends from the flattened Kuiper Belt to the spherical Oort Cloud. We
will quantify the effect of the Scattered Disc by calculating the quadrupole moments of its thermal emission. This will exist
irregardless of any externally induced asphericity in the Oort Cloud.

An increasing body of evidence now suggests that the Solar System did not form in isolation, but in a stellar cluster which subsequently 
evaporated \citep{Adams01,Hester04}. This is not surprising since most low mass stars are observed to form in clusters that subsequently 
evaporate when the interstellar gas, which is required to gravitational bind the cluster, is expelled from the cluster by either stellar 
winds or supernova feedback \citep{Lada03}. Beyond this general trend, anomalous abundance ratios are observed in the Solar System that would 
indicate that a Type II supernova occur somewhere in the vicinity of the Solar System during its infancy \citep{Looney06}. With a higher 
number density of stars in the cluster the rate of external perturbations would have been higher and the spherical inner edge of the Oort 
Cloud would have extended to smaller semi-major axes.

In this paper we will consider the dynamical effect that a star passing within $\lesssim 1 \pc$ of the sun would have had on the Oort 
Cloud. And we calculate the signatures that would be produced in the CMB frequency spectrum. The outline of this paper is as follows. 
In \S \ref{sec:signal} we describe how a non-sherical distribution of Oort Cloud objects may affect the measured brightness 
distribution of the CMB on the sky. In \S \ref{sec:poisson} we calculate the variance of the Oort Cloud signal due to Poisson fluctuations, 
whereas in \S \ref{sec:ic} we describe the initial distribution of Oort Cloud objects.  In \S \ref{sec:calc} we outline of method of 
calculation the perturbations in the orbital elements of the Oort Cloud objects, and in \S \ref{sec:impact} we calculate the distribution 
of impact parameters of stellar perturbers. In \S \ref{sec:virial} we derive an formula based on the collisionless Boltzmann equation that 
allows us to analytically estimate the resulting quadrupole distortion from a perturbation. Our results are discussed in \S \ref{sec:results}. 
In \S \ref{sec:bayes} we develop a Bayesian estimator that can be used to constrain the parameters of a stellar encounter from the observed 
CMB data. Finally \S \ref{sec:concl} summarizes our main conclusions, Appendix \ref{sec:quad} derives relationships between various measures 
the quadrupole moment and Appendix \ref{sec:noise} describes the model of instrumental noise for the Planck satellite.

\section{Calculation}
\label{sec:signal}

We start by calculating how the non-sphericity of the Oort Cloud affects the measured angular distribution of the CMB intensity. The 
fluctuation in the CMB frequency spectrum along a given direction
\begin{equation}\label{eq:delta_I}
\delta I_{\nu}(\hat{\bm n}) \equiv I_{\nu}(\hat{\bm n}) - B_{\nu}(\bar{T}_{\rm CMB}),
\end{equation}
can be expressed as
\begin{align}
\label{eq:delta_int}
\delta I_{\nu}(\hat{\bm n}) =& \Delta T(\hat{\bm n})\frac{\partial B_{\nu}(T)}{\partial T}  \\ 
& + f_A \int dx_i p(x_i) \frac{1}{D(x_i)^2} [B_{\nu}(T_{\rm Oort}(x_i))- B_{\nu}(\bar{T}_{\rm CMB})], \nonumber
\end{align}
where $p(x_i)$ is the probability distribution of orbital elements describing the Keplerian orbits of the Oort Cloud 
objects around the Sun. The first term on the right is due to the extragalactic CMB temperature anisotropies. The CMB 
signal that comes from the surface of last scattering ($z=1000$) has a blackbody frequency distribution that is parameterized 
by slightly different temperatures in different directions \citep{Dodelson03}. Once we have subtracted off the frequency spectrum
corresponding to the average CMB temperature, Eq. (\ref{eq:delta_I}), the CMB temperature anisotropies induce some 
spectral distortion since we have removed a blackbody at an incorrect temperature for that particular pixel. The second 
term is caused by the Oort Cloud objects in the outer Solar System as described in \S \ref{sec:intro}, where $D$ is the distance
of the Oort Cloud object from the Sun. The absolute covering fraction is
\begin{equation}
\label{eq:opt}
f_A = \int dM \frac{R^2}{4} \frac{dn}{dM}(M),
\end{equation}
where $R$ is the radius of the Oort Cloud object. The absolute covering fraction is related to the optical depth as $\tau = f_A/D^2$. 
Since the individual objects are optically thick, but sparsely fill the telescope beam function with which the CMB sky has been observed, 
the optical depth is simply related to the covering fraction of Oort Cloud objects. We add the term 'absolute' because $f_A$ is the 
covering fraction if all Oort Cloud objects were placed at a standard distance. We have implicitly assumed that the physical size 
and the orbital elements of the Oort Cloud objects are uncorrelated. This assumption allows us
to decouple the probability distribution of the orbital elements used in Eq. (\ref{eq:delta_int}) from the size distribution,
defined in Eq. (\ref{eq:mass_func}) and used to calculate the absolute covering fraction. This is a reasonable assumption 
since the comets behave as test particles with respect to Sun and the perturbing star\footnote{Note that it is not obvious 
that the initial distribution of orbital elements should be independent of the object's mass because non-gravitational forces, 
such as gas drag \citep{Weidenschilling77,Rafikov04} and Pointying-Robertson drag \citep{Burns79}, may have been important 
when the gas giant planets were scattering the comets into highly eccentric orbits. Also, collisional effects, which are more 
important for small objects, may alter the formation of the Oort Cloud \citep{Stern01,Charnoz03,Charnoz06}. These are 
issues that go beyond the scope of this paper.}. For the internal density of the Oort Cloud objects we adopt a value 
of $\rho = 1 \g \cm^{-3}$.

In order to relate the mean optical depth to a total mass of the inner Oort Cloud, the differential mass function must 
be specified. We model it as a broken power-law\footnote{Typically the differential mass function is expressed in terms of 
comet radius, not mass. The power-law index with respect to radius $\alpha_{\rm radius}$ is related to the values used in
this paper as $\alpha_{\rm mass} = (\alpha_{\rm radius}+2)/3$.}
\begin{align}\label{eq:mass_func}
\frac{dn}{dM}(M) = & A M^{-\alpha}, & M_{\rm min} < M < M_{\rm br}, \\
\frac{dn}{dM}(M) = & A M_{\rm br}^{-\alpha+\beta} M^{-\beta}, & M_{\rm br} < M < M_{\rm max}. \nonumber
\end{align}
The mass distribution yields the total mass through the integral
\begin{equation}
M_{\rm tot} = \int dM M \frac{dn}{dM}(M).
\end{equation}
The empirical values of the above power-law indices are not known. For the Kuiper Belt, the slope of the high mass distribution 
has been restricted to the range $\beta = 2 \pm 0.2$ \citep{Bernstein04}. There are no strong observational constraints on the 
slope of the low mass distribution, $\alpha$; however, it can be calculated through the assumption of collisional equilibrium 
\citep{Pan05}. The exact value of the slope then depends on the assumed material strength of the Kuiper Belt objects. Similiar 
constraints on the inner Oort Cloud mass distribution does not exist. Since collisions have not been numerous enough to affect
the slope of the high mass distribution, it is reasonable to adopt the observed value of the slope of planetesimals in the 
proto-planetary disc and use it for the distribution of inner Oort Cloud objects. We will adopt $\beta = 2.16$ for the calculations 
in this paper. The collisional grinding of the Kuiper Belt objects, which justifies the above assumption of collisional equilibrium, 
is believed to have occurred amongst the objects at their current location. The collision timescales in the Oort Cloud are
much longer, so a state of collisional equilibrium cannot have been reach if the objects were always at their present locations. 
Fortunately collisions are believed to have been frequent during the formation phase of the Oort Cloud while the objects were still 
in the planetary region of the Solar System \citep{Stern01,Charnoz03,Charnoz06}. Again we will adopt the Kuiper Belt value of $\alpha = 1.83$. 
Our choice of this value is conservative, since the primordial slope was likely steeper, providing even better constraints than we 
will derive.

We assume $M_{\rm max} \equiv M_{\rm Sedna} = 6 \times 10^{25} \mbox{ g}$ \citep{Brown07}, as the upper mass limit. Our results are insensitive 
to changes in the upper mass limit for $\beta > 2$. We will take the minimum mass to correspond to objects of radius $1 \mm$. This quantity depends 
on the detailed formation history of the Oort Cloud and is, therefore difficult to calculate from first principles. The minimum object mass is 
partially degenerate with the total mass of the outer Solar System which is also unknown, so for the purposes of this paper we will fix the minimum 
radius at $1 \mm$. However, we note that the efficiency of blackbody emission is suppressed when the radiation wavelength is larger than the size 
of object emitting the radiation due to Kirchoff's Law \citep{Greenberg78,Backman93,Landau99}. 
It should be emphasized that we only need our adopted model of the size 
distribution to convert a total mass of the Oort Cloud into an optical depth; if we simply choose a value for the optical depth then the rest of 
this paper would be independent of our assumed size distribution.

For the location of the break in our mass model, we assume $M_{\rm br} = 4 \times 10^{15} \mbox{ g}$ (corresponding to $R_{\rm br} = 1 \mbox{ km}$). 
This value is established by requiring that objects of this size were likely to have undergone a collision during the time period of Oort Cloud's 
formation. The formation of the Oort Cloud most likely was a continuous process so it is difficult to precisely define its formation time. 
In simulations this time period can roughly be estimated as $\Delta t_{\rm Oort} = 200 \Myr$ based on a visual inspection (see Fig. 8 of 
\citet{Dones04}). The collision rate of objects of size $R$ in a protoplanetary disc is \citep{Goldreich04}
\begin{equation}
\Gamma = \frac{\sigma \Omega}{\rho R},
\end{equation} 
where $\sigma$ is the surface mass density of planetesimals, $\Omega$ is the Keplerian velocity and $\rho$ is the planetesimal density. We adopt 
minimum mass solar nebula value for the surface density of condensates $\sigma = (20 \au/ a)^{3/2} \gcm$ \citep{Goldreich04}.
Now the break radius can be calculated as
\begin{align}
R_{\rm br} &= \frac{\sigma \Omega \Delta t_{\rm Oort}}{\rho} ,\\
&= \frac{1 \km}{(a/20 \au)^3}.
\end{align}
A full simulation would be required to precisely determine this quantity as well as its radial dependence within the Oort Cloud. This is roughly 
consistent with the HST results of \citet{Bernstein04} for the Kuiper Belt. Our results are insensitive to the precise choice of these values 
since they are dominated by the smallest objects which have the largest surface area to volume ratio.

The total mass in the Oort Cloud is calculated by requiring there to be $10 M_{\oplus}$ in objects $R \ge 1 \km$ in the outer Oort Cloud 
\citep{Weissman90}. This figure is related to the observed number of new long period comets and implies there is approximately $10^{13}$ comets 
present in the outer Oort Cloud. We use our adopted models to relate this number of comets to the total mass in the inner Oort Cloud. For simplicity 
we assume equal amounts of mass in the inner and outer Oort Clouds. Simulations of the formation of the Oort Cloud have implied that the inner 
Oort Cloud is more massive \citep{Duncan87}, while others have implied that the outer Oort Cloud is more massive \citep{Dones07}. This uncertainity, 
as well as the uncertainity in other aspects of our model, should be remembered when interpreting our results. In Table \ref{qaz} we present 
slightly different models and calculate the total number of objects, the total mass and the effective optical depths of these models. We quote the 
absolute covering fraction $(f_A)$ divided by the mean squared distance
\begin{equation}
\frac{1}{\langle a^2 \rangle} = \int da P(a) \frac{1}{a^2}.
\end{equation}
The distribution of semi-major axes and the limits of integration are described in \S \ref{sec:ic}. For definitiveness we adopt $P(a) \propto a^{-3}$;
$a_{\rm max} = 10^4 \au$ and $a_{\rm min} = 10^3 \au$ for our calculations in Table \ref{qaz}. These parameter choices lead to the value
\begin{equation}
\frac{1}{\langle a^2 \rangle} = 5.05 \times 10^{-7} \au^{-2}.
\end{equation}

\begin{table*}
\begin{center}
\begin{tabular}{cccccc}
\hline 
\hline
\hline 
$R_{\rm min} (\cm)$ & $R_{\rm br} (\km)$ & $\alpha$ & $N_{\rm Oort}$ & $M_{\rm total} (M_{\oplus})$ & $\tau = f_A/\langle a^2 \rangle$ \\
\hline 
\hline 
0.1 & 10 & 1.83 & $6.2 \times 10^{28}$ & $1184$ & $1.8 \times 10^{-6}$ \\
0.1 & 1 & 1.83 & $1.0 \times 10^{29}$ & $590$ & $2.9 \times 10^{-6}$ \\
0.1 & 10 & 2.0 & $7.1 \times 10^{31}$ & $2565$ & $1.2 \times 10^{-3}$ \\
0.01 & 10 & 1.83 & $8.1 \times 10^{31}$ & $1194$ & $2.1 \times 10^{-5}$ \\
\hline
\hline
\hline 
\end{tabular}\end{center}
\caption{\label{qaz} Total number ($N_{\rm Oort}$), total mass ($M_{\rm total}$) and absolute covering fraction ($f_A$) of 
Oort Cloud objects for the adopted model in this paper (the first row) and slight changes to the model. Also shown is the
optical depth ($\tau$) for each model. We calculate these quantities by requiring that $50 M_{\oplus}$ of comets ($R \ge 1 \km$) exist.}
\end{table*}

Next we calculate the microwave anisotropies induced by the thermal emission of the inner Oort Cloud. We consider the real-space quadrupole 
moments instead of the more conventional spherical harmonic moments, because it is easier to interpret the results in real space. These moments
are defined as
\begin{equation}\label{eq:real_quad}
Q_{i j}(\nu) = \int d^2\hat{\bm n}~(\hat{\bm n}_i \hat{\bm n}_j - \frac{1}{3}\delta_{ij}) \frac{\delta I_{\nu}(\hat{\bm n})}{I_{\rm CMB}},
\end{equation}
where $I_{\rm CMB}= B_{\nu}(T_{\rm CMB})$ is the homogeneous intensity of the CMB\footnote{We should note that if the Oort 
Cloud was intrinsically spherical in a heliocentric coordinate system and the directional vectors used to measure the quadrupole
moments, $\hat{\bm n}$ in Eq. (\ref{eq:real_quad}), were defined in a geocentric coordinate system, then a small quadrupole 
moment would be observed. This quadrupole moments is roughly $Q \sim (r_{\oplus}/r_{\rm Oort})^2 \delta I \sim 10^{-6} \delta I$ 
and therefore quite small.}. We also quote our results as temperature fluctuations, which are more commonly used, in \S \ref{sec:concl}.

There are five independent quadrupole moments because the quadrupole tensor is symmetric and traceless. The Oort Cloud quadrupole moments 
are also frequency dependent. The standard CMB moments are frequency independent\footnote{This statement is only true when the primordial
CMB moments are expressed in terms on temperature, not intensity, flucuations. The Oort Cloud moments expressed as either temperature or
intensity fluctuations are frequency dependent.} because the usual sources of the cosmological temperature anisotropy, gravitational redshift 
and Thomson scattering, do not depend on the frequency of the scattered radiation \citep{Dodelson03}. Of course, we can relate the real space 
quadrupole moments to the spherical harmonic moments. These relationships and the conversion between different measures of the quadrupole 
are discussed in Appendix \ref{sec:quad}.

\subsection{Poisson Fluctuations}
\label{sec:poisson}

The flux emitted by a single unresolved Oort Cloud object scales as
\begin{equation}\label{eq:scale}
f_{\nu} \propto \frac{1}{D^2} B_{\nu}[T_{\rm Oort}(D)],
\end{equation}
which in the Rayleigh-Jean's portion of the spectrum scales as $f_{\nu} \propto D^{-5/2}$. 
At higher frequencies in the Wien portion of the spectrum the scaling has a steeper dependence 
on distance $D$. Consequently, even when we calculate the time average of the quadrupole moments
\begin{align}\label{eq:scale1}
\langle f_{\nu}(\hat{\bm n}) \rangle_T  &= \frac{1}{P} \int dt f_{\nu}(\hat{\bm n}), \\
&\propto \int df D^2 f_{\nu}(\hat{\bm n}),
\end{align}
the integrand still scales as an inverse power of $D$. In the above
equation we have used Kepler's second law\footnote{This states that a
vector joining two masses will sweep out equal areas in equal times, which
is equivalent to the conservation of the specific angular momentum, $h =
r^2 \dot{f}$.} to change integration variables from time ($t$) to true
anomaly ($f$). In Eq. (\ref{eq:scale1}) $P$ is the orbital period of the Oort Cloud object.
Therefore, an individual Oort Cloud object will produce the
largest amount of signal per unit time when it is closest to perihelion and
equivalently the comets that happen to be nearest to their perihelia will
contribute the greatest to the overall signal per unit surface area. Since
the comet's velocity is largest close to perihelion there will be
fluctuations in the direction and amplitude of the quadrupole. These
fluctuations will become larger at high frequencies where the signal's
scaling with inverse distance is even stronger. The signal in a given
direction is the average signal of all Oort Cloud objects weighted by their
apparent angular size. If the relatively few nearby objects that appear
large, have a higher temperature and move quickly, dominate the signal over
the more numerous farther away objects, then the Poisson fluctuations will
be larger.

We adopted a continuum description of the Oort Cloud object masses and orbital elements, which implicitly assumed 
an infinite number of objects, when calculating the induced CMB spectral distortions in \S \ref{sec:signal}. In 
reality there are only a finite number of objects along any given line-of-sight and the signal will be dominated 
by the smaller number of objects that are closest to perihelion, so Poisson fluctuations will produce statistical 
fluctuations. These fluctuations will be estimated in this section.  

The surface brightness in a given direction, which is now a random variable, can be expressed as a sum over the 
contribution from any object which lies along the line-of-sight,
\begin{align}\label{eq:poisson}
\delta I_{\nu}(\hat{\bm n}) = & \frac{1}{N} \sum_{i=1}^N \theta^{(2)}(R_i/r_i - |\hat{\bm n} - \hat{n}_i|) \\
& \times [B_{\nu}(T^i_{\rm Oort}))- B_{\nu}(\bar{T}_{\rm cmb})], \nonumber
\end{align}
here $\hat{n}_i$ is the position on the sky of the Oort Cloud object's centroid and the $R_i$ is the Oort Cloud 
object's radius and $r_i$ its radial distance. The $\theta$-function simply requires that the direction of interest 
intersects the relevant Oort Cloud object. Here N is the total number of objects in the sample that we are 
considering. We can directly calculate the Poisson fluctuations
\begin{align}
\langle \delta I_{\nu}(\hat{\bm n}_1) & \delta I_{\nu}(\hat{\bm n}_2) \rangle =  \frac{1}{N^2} \sum_{i=1}^N \langle
\theta^{(2)}(R_i/D_i - |\hat{\bm n}_1 - \hat{n}_i|) \\
&\times \theta^{(2)}(R_i/D_i - |\hat{\bm n}_2 - \hat{n}_i|) 
[B_{\nu}(T^i_{\rm Oort}) - B_{\nu}(\bar{T}_{\rm cmb})]^2 \rangle.\nonumber
\end{align}
The Oort Cloud objects subtend a finite angle on the sky and Poisson fluctuations will cause CMB fluctuations if
$\hat{\bm n}_1$ and $\hat{\bm n}_2$ intersect the same object. The expectation value is taken over the assumed 
distribution of orbital elements and masses. 

We are interested in the intensity fluctuation averaged over some angular aperature, which we take to be a top-hat
of angular size $\bar{\theta}$,
\begin{equation}
\bar{\delta I}_{\theta} = \int \frac{d^2\hat{\bm n}}{\bar{\theta}^2} \delta I_{\nu}(\hat{\bm n}). 
\end{equation}
The power induced in the CMB by the Poisson fluctuations can be related the variance of the aperature averaged 
intensity fluctuations as 
\begin{equation}
\frac{\ell^2 C_{\ell}}{2\pi} = \langle \left( \frac{\bar{\delta I}_{\nu}}{B_{\nu}(\bar{T}_{\rm cmb})} \right)^2 \rangle,
\end{equation}
where smoothing was done on the angular scale $\theta = 2\pi/\ell$.

At a fixed time the average over the initial time of perihelon is equivalent to performing a time average over 
the orbital period which can be related to an average over the true anomaly. After we perform these averages we 
find that the power spectrum produced by Poisson fluctuations is
\begin{align}
C_{\ell} = & \frac{1}{2\pi} \frac{1}{N} \int dM da de d\Theta P(\Theta) 
P(a) P(e) P(M) \\
&\times \pi \left(\frac{R}{D}\right)^2 \left[ \frac{B_{\nu}(T_{\rm Oort}(D))}{B_{\nu}(\bar{T}_{\rm cmb})} - 1 \right]^2. \nonumber
\end{align}
The distributions of orbital elements and mass are given in Eqs. \ref{eq:ic} \& \ref{eq:mass_func}, respectively. 
The integrals over the three orientation angles are unity. Although we will apply this formula to the quadrupole 
$(\ell=2)$, it is applicable to the power spectrum on any angular scale.

There are two distinct reasons that we are interested in Poisson fluctuations. The first involves understanding the 
accuracy of our numerical calculations, described in \S \ref{sec:calc}. We will perform our calculation with $10^3$ 
realizations of $10^6$ test particles, but due to the stronger weighting of the test particles closest to perihelia, 
the effective number of test particles is reduced and the fractional fluctuation amplitude increased. The second is 
related to the statistical significance of any observed CMB spectral distortion. As Oort Cloud objects move along 
their orbits the closest objects, and therefore the objects that produce the largest signal, will change. These 
Poisson fluctuations could potentially limit the statistical significance and therefore usefulness of any measurement 
of CMB spectral distortions. The results of our calculations are shown in Table \ref{table:poisson} for the fractional 
fluctuations relevant to both our code and the real Oort Cloud.

\begin{table}
\begin{center}
\begin{tabular}{ccc}
\hline 
\hline
\hline 
$\nu$ & $C_2^{\rm OC}$ & $C_2^{\rm code}$ \\
\hline 
\hline 
$30 \GHz$ & $1.3 \times 10^{-32}$ & $1.3 \times 10^{-14}$ \\
$100 \GHz$ & $5.2 \times 10^{-32}$ & $5.2 \times 10^{-14}$ \\
$217 \GHz$ & $7.4 \times 10^{-31}$ & $7.4 \times 10^{-13}$ \\
$353 \GHz$ & $2.5 \times 10^{-29}$ & $2.5 \times 10^{-11}$ \\
$545 \GHz$ & $5.4 \times 10^{-27}$ & $5.4 \times 10^{-9}$ \\
\hline
\hline
\hline 
\end{tabular}\end{center}
\caption{\label{table:poisson}Fractional Poisson fluctuations in the number of Oort Cloud objects at different 
frequencies. The fractional fluctuation relevant for the calculations of our code $(C_2^{\rm code})$ and the real
Oort Cloud $(C_2^{\rm OC})$ are shown.}
\end{table}

We conclude that statistical fluctuations will not be important when interpreting the observational data if our 
model of the size distribution of Oort Cloud objects is reasonable. These fluctuations will be much smaller than 
standard experimental noise. However, the statistical fluctuations present in our numerical calculations, which 
contain significantly fewer test particles than the true number of Oort Cloud objects, will be considerably larger 
and comparable to the magnitude of the effect that we are interested in. Therefore we will calculate the distribution 
of quadrupole moments by calculating the quadrupole at $10^3$ different times. From this distribution we can estimate 
the magnitude of the perturbation on the shape of the Oort Cloud, as well as, the effect of Poisson fluctuations on 
the resultant CMB spectral distortions.

\subsection{Initial Conditions}
\label{sec:ic}

The comets in the Oort Cloud, in the absense of external perturbations, effectively act as test particles that move on Keplerian 
orbits around the Sun. The hypothetical stellar encounter simply perturbs the comets onto Keplerian orbits with new orbital 
elements\footnote{We also do allow the comets to be ejected from the Solar System if the final eccentricity if greater than one.}.  
The orbits are completely described by six orbital elements -- (i) the semi-major axis ($a$); (ii) the eccentricity ($e$); 
(iii) the inclination of the orbital plane ($i$); (iv) the longitude of perihelion ($\omega$); (v) the longitude of the 
ascending node ($\Omega$); and (vi) the time of perihelion\footnote{Ordinarily $\tau$ is used to designate the time of 
perihelion passage, we chose a different symbol to prevent confusion with the use of $\tau$ for the optical depth.} ($\Theta$). 
The inclination angle, the longitude of perihelion and the longitude of the ascending node, are rotation angles used to convert 
the coordinate system of invariant plane (which is different for every comet) to a standard reference coordinate system. With 
these orbital elements the motion of the comet around the Sun is completely specified.

We set the initial conditions of the Oort Cloud objects by randomly sampling the uniform distributions
\begin{align}\label{eq:ic}
0.3 &\le e \le 1 - \frac{50 \au}{a}, \\
-1 &\le \cos{I} \le 1, \\
0 &\le \omega \le 2\pi, \\
0 &\le \Omega \le 2\pi, \\
0 &\le \Theta \le P,
\end{align}
where $P = 2\pi a^{3/2}/(G M_{\odot})^{1/2}$ is the test particle's orbital period. The maximum eccentricity is determined by 
the condition that the comet's perihelion lies far enough away from Neptune that it becomes dynamically decoupled. This condition 
roughly requires $r_p = a(1-e) \ge 50 \au$, so the maximum eccentricity depends on on the semi-major axis of the orbit. Simulations 
of the formation of the Oort Cloud have found that the probability distribution of semi-major axes has a power-law form 
$P(a) \propto a^{-\gamma}$, but have disagreed on the value of its index $\gamma$. Originally it was stated by \citet{Duncan87} 
that $\gamma \sim 3.5$, but updated simulations with more realistic initial conditions for the comets have found a shallower 
value, $\gamma \sim 1.5$ \citep{Dones07}. When we present numerical results in \S \ref{sec:results} we adopt the value of $\gamma = 3$.
We will assume that the inner edge of the Oort Cloud is $a = 10^3 \au$. We take the inner edge to be semi-major axis beyond 
which the Oort Cloud becomes roughly spherical. The Scattered Disc lies inside of this radius.
And we pick the outer edge to lie at $a = 10^4 \au$. We then randomly choose an initial time 
$0 \le t_i \le P$ to calculate the initial position in order to evaluate the above perturbation equations. Then we choose another 
random final time $0 \le t_f \le P$ in order to evaluate the final positions of the comets. 

The initial conditions of the Oort Cloud before the stellar encounter
depend on the rate at which the comets are scattered to high inclination
angles\footnote{The initial distribution of $\omega$ and $\Omega$ should
have been roughly uniform even in the proto-planetary disc.}. At smaller
distances, the isotropization processes are less efficient and the cloud
will be smoothly compressed into a torus that extends out of the ecliptic
plane. The initial conditions depend on the efficiency of these relaxation
processes. These processes include the effects from passing stars and
molecular clouds, the tides of the Galactic disc, and the potential of the
parent molecular cloud inside of which the Sun had formed. Placing a large
amount of mass into the non-spherical Scattered Disc or
any primordial asphericity in the Oort cloud will only increase the
anisotropic spectral distortions. The intensity fluctuations induced by the
Scattered Disc estimated in this paper can be viewed as a lower limit for
the signal. 

\begin{figure}
\centering
\includegraphics[width=8cm,height=8cm]{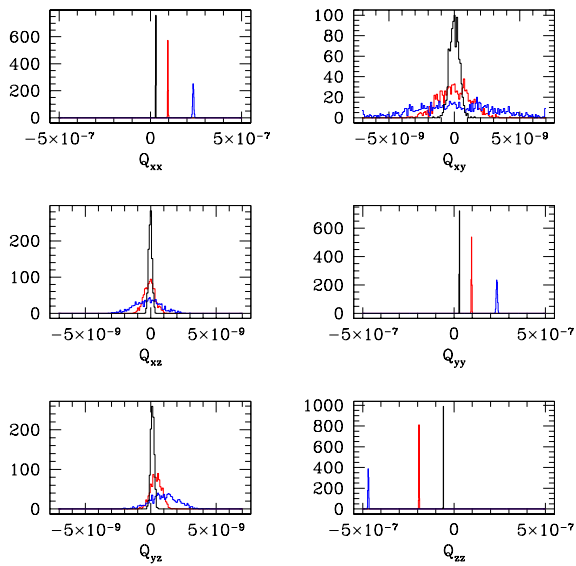}
\caption{\label{fig:scatt} Histograms of the six quadrupole moments at three different frequencies 
$\nu = 30 \GHz$ (black), $\nu = 143 \GHz$ (red) and $\nu = 217 \GHz$ (blue). The mixed quadrupole 
moments all vanish within Poisson fluctuations and the diagonal moments have are nonzero at a 
statistically significant level.}
\end{figure}

We estimate of the influence of the initial conditions by calculating the
expected quadrupole signal from the Scattered Disc.  We assume that the
total mass is $0.1 M_{\oplus}$ and the semi-major axes lie between $10^2
\au$ and $10^3 \au$. Most importantly we choose the disc's opening angle to
be $30$ degrees. The six quadrupole moments at three frequencies, $\nu = 30
\GHz$ (black), $\nu = 143 \GHz$ (red) and $\nu = 217 \GHz$ (blue), are
shown in Fig. \ref{fig:scatt}. The mixed quadrupole moments, such as
$Q_{xy}$, all vanish within Poisson fluctuations. As expected Poisson
fluctuations become larger at higher frequencies.  The diagonal quadrupole
moments, such as $Q_{xx}$, do not vanish. The $Q_{zz}$ moment does not
vanish because the disc is compressed into the Ecliptic Plane. The
traceless condition, $Q_{xx} + Q_{yy} + Q_{zz} = 0$, forces the other two
diagonal moments to be non-zero. Since the scattered disc is azimuthally
symmetric by construction, $Q_{xx}$ and $Q_{yy}$ are forced to be equal.

\subsection{Orbital Perturbation}
\label{sec:calc}

The external perturbation from a passing star acts both on the Oort cloud as well as the Sun which are bound together, 
and so only the relative (tidal) acceleration between the comet and the Sun is relevant
\begin{align}
\Delta {\bm a} \equiv& {\bm a}_{\rm Oort} - {\bm a}_{\odot}, \nonumber \\
&= \frac{G M_p ({\bm b} - {\bm r})}{|{\bm b}- {\bm r}|^3} - \frac{G M_p {\bm b}}{b^3},
\end{align}
where $M_p$ is the mass of the perturber and ${\bm b}$ is the vector from the Sun to the perturber. 

Kepler's third law, $T \propto a^{3/2}$, implies that an object on an orbit with $a = 1000 \au$ has an orbital period, $T = 10^{12} \s$. 
If the impact parameter of the perturber is $b \approx 2000 \au = 0.01 \pc$ and it is traveling at a speed of $v = 30 \km/\s$, then the 
effective duration of the impulse would be $T_{\rm pert} \lesssim 2 b/v = 2 \times 10^{10} \s$. The orbital period is typically longer 
than the duration of the impulse, and so we will use the impulse approximation for simplicity. Numerical experiments have found that the 
validity of the impulse approximation even extends to the regime where $T \approx 0.2 \times T_{\rm pert}$ \citep{Aguilar85}. For test 
particles with larger semi-major axes the accuracy of the impulse approximation becomes increasingly better.

Clearly, a passage through the Oort Cloud would strongly perturb the comets that lie nearly along the perturber's trajectory \citep{Weissman90}. 
Such an effect is believed to be responsible for the comet showers that occasionally take place (as the strongly perturbed comets either enter 
the inner Solar System or get ejected from the Oort Cloud), but will these strongly perturbed Oort Cloud objects will have little effect on 
results because they are assumed to have been removed from the Oort Cloud. In our calculation will assume that all test particles that are more
strongly attracted to the perturber than the Sun are ejected. Initially this has a large effect on the angular structure of the Oort Cloud, 
however the effect is quickly diminished due to phase mixing. After a test particle is ejected from the Solar System a hole is left in the 
Oort Cloud that orbits around the sun with the orbital parameters of the original test particle. This results in an underdensity of test particles 
that would produce anisotropy in the same manner as an overdensity would. However the large spread in orbital periods, due to the large spread 
in semi-major axes, means that the signal vanishes over a timescale corresponding to the orbital period of the test particles. Of course, the mean 
signal will remain the same as this only depends on the total covering fraction of Oort Cloud objects. However the quadrupole, or any 
angular moments, will vanish as the holes left by the ejected test particles lose spatial coherence.

In the impulse approximation, only the component of the force perpendicular to the perturber's motion remains after its path is integrated 
over \citep{Binney87}. Therefore we must apply the projection operator 
\begin{equation}\label{eq:proj}
{\bm T} = {\bm I} - \hat{\bm v} \hat{\bm v},
\end{equation}
to the tidal force, where ${\bm v}$ is the velocity of the perturber. The components of the projected tidal force, written in a coordinate basis 
attached to the comet, are
\begin{align}
R &= \hat{\bm e}_R \cdot {\bm T} \cdot \Delta {\bm a}, \\
T &= \hat{\bm e}_T \cdot {\bm T} \cdot \Delta {\bm a}, \\
N &= \hat{\bm e}_N \cdot {\bm T} \cdot \Delta {\bm a}, 
\end{align}
here $\hat{\bm e}_R$ is the radial basis vector, $\hat{\bm e}_T$ is the tangential basis vector and $\hat{\bm e}_N$ is the normal basis vector.
The Lagrange equations describe how the perturbation changes the orbital elements \citep{Danby62,Murray00}
\begin{align}\label{eq:pert}
\frac{da}{dt}=& \frac{2a^{3/2}}{\sqrt{\mu(1-e^2)}}[R~e \sin{f} + T(1+e\sin{f})], \\
\frac{de}{dt}=& \sqrt{\frac{a(1-e^2)}{\mu}}\left[R \sin{f} + T(\cos{f} + \frac{e+\cos{f}}{1+e\cos{f}})\right],\\
\frac{di}{dt}= & \sqrt{\frac{a(1-e^2)}{\mu}}\frac{N \cos{(\omega+f)}}{1 + e\cos{f}}, \\
\frac{d\Omega}{dt}=& \sqrt{\frac{a(1-e^2)}{\mu}} \frac{N \sin{(\omega+f)}}{\sin{i}(1+e\cos{f})}, \\
\frac{d\omega}{dt}=& \sqrt{\frac{a(1-e^2)}{e^2\mu}}[-R\cos{f} + T\sin{f}\frac{2+e\cos{f}}{1+e\cos{f}}] \nonumber \\
&-\frac{d\Omega}{dt}\cos{i} , \\
\frac{d\Theta}{dt} =& ~3 R (\Theta-t) \sqrt{\frac{a}{\mu(1-e^2)}}e\sin{f} \\
&+ R\frac{a^2(1-e^2)}{\mu} \nonumber
(\frac{2}{1+e\cos{f}} - \frac{\cos{f}}{e}) \\ \nonumber
+& ~3 T (\Theta-t) \sqrt{\frac{a}{\mu(1-e^2)}}e\sin{f} \\ \nonumber
&+T \frac{a^2(1-e^2)}{e \mu} \frac{\sin{f}(2+e\cos{f})}{1+e\cos{f}}, \nonumber
 \end{align}
where $\mu = G M_{\odot}$. 

Within the impulse approximation, we assume that the internal configuration of the solar system 
is frozen during the close encounter with the perturber; it is this assumption that causes the 
force parallel to the perturber's velocity to exact cancel. Therefore the change in the orbital
element, x, can be expressed as
\begin{equation}
\Delta x \approx \frac{dx}{dt} \frac{2b}{v}.
\end{equation}
The new distribution of orbital elements will be solely due to the addition of velocity perturbations, 
because all previous comets with the given elements will have been perturbed to other orbits
\begin{equation}
 p(\{ x_i\}) = \int p_0(x_i - \Delta x_i) D(\Delta x_i)
\end{equation}

In the test particle limit, the mass of the comet should not affect how it is perturbed by the stellar 
companion. Therefore the distribution function of the Oort Could comets should be factorizable into the
distribution of comet orbital elements and the mass function of comets
\begin{equation}\label{eq:dist}
f_{\rm Oort} = \frac{dn}{dM}(M) \times p(\{x_i\}).
\end{equation}

We will use the orbital element distribution function in Eq. (\ref{eq:dist}) to calculate the CMB intensity 
fluctuation using Eq. (\ref{eq:delta_int}). In principle the orbital element distribution function can be used 
to construct a distribution function parameterized by the Keplerian integrals of motion. Since the distribution
function is six dimensional and we are interested in observable signatures in the CMB, we will not focus on this 
aspect of the problem. However we will mention a condition that the distribution function must satisfied if an 
angular signal will be produced. For a quadrupole, or any angular moment, intensity fluctuation to be non-zero, 
the distribution of Euler angles ($\Omega,\omega,i$) must be correlated with the distribution of semi-major axes 
and eccentricities. In our calculations we initially assume that all the orbital elements are uncorrelated and
that the Oort Cloud is statistically isotropic. The stellar perturbation induces the necessary correlations to 
produce a statistically significant signal. 

\subsection{Likely Encounter Parameters}
\label{sec:impact}

In addition to determining the initial conditions of the Oort Cloud objects we need to specify the properties 
of the perturber. Observations of local stars can identify their impact parameter with respect to the Solar System 
only up to $ 10 \Myr$ in the past \citep{Garcia99, Garcia01}. Of course, we can use the initial mass function 
of stars and their velocity dispersion with respect to the local standard of rest to make statistical statements 
about the likely properties of the perturber, however for a single perturber that gets to the closest approach 
and has the biggest effect, such statistical statements are not very useful. In \S \ref{sec:bayes} we will discuss 
a Bayesian method that allow us to use a detected signal to statistically constrain the properties of the perturber. 
Given a number density and a relative velocity with respect to the Sun the minimum impact parameter of the perturber, 
$R_{min}$, can be estimated by requiring that the probability of encounter during some time period, $\Delta t$, is 
of order unity
\begin{equation}\label{eq:impact}
P = \pi R^2 n_{\rm stars} v \Delta t \approx 1.
\end{equation} 

The environment of the Solar System might have been quite different in the past than it is today. As mentioned 
in \S \ref{sec:intro}, the Sun might have formed in a stellar cluster and so the number density of neighboring
stars may have been much higher. Also as the Solar System undergoes epicyclic and vertical oscillations with respect 
to the Galactic disc or passes through spiral arms, the environment also changes. The closest encounter probabilities  
can be adjusted to account for the changing Solar environment.

A variety of relaxation processes -- passing stars and giant molecular clouds \citep{Duncan87,Hut85}, precession due 
to the potential of the Galactic disc \citep{Heisler86} and the self gravity of the Oort Cloud \citep{Tremaine05} -- 
can isotropize the orbits of objects in Oort Cloud and reduce the non-sphericity produced by the stellar encounter. 
Of course the possibility of eliminating these signatures does not imply that the Oort Cloud will be spherical today, 
but that it will possess the non-sphericity induced by the closest stellar encounter that occurred at a time less 
than the appropriate relaxation timescale. The self-gravity of the Oort Cloud was determined to produce the largest 
effect; the calculated timescale was $1 \Gyr$ \citep{Tremaine05}. More calculations are needed to determine the effect 
on nearly parabolic orbits, which spend considerable amounts of time at different distances and therefore experience 
different gravitational perturbations. We will leave a detailed analysis of these relaxation processes to future work.
Therefore we will consider the minimum impact parameters for the current Solar neighborhood with a variety of 
time intervals -- $\Delta t = 0.5, 1, 4 \Gyr$. These models are described in Table \ref{table:impact}.

\begin{table*}
\begin{center}
\begin{tabular}{ccccc}
\hline 
\hline
\hline 
Model & $n (\ipc)$ & $v_{rel} (\kms)$ & $\Delta t (\Gyr)$ & $R_{min} (\au)$ \\
\hline 
\hline 
Present Neighborhood - A & 0.1 & 30 & 4 & 1100 \\
Present Neighborhood - B & 0.1 & 30 & 1 & 2100 \\
Present Neighborhood - C & 0.1 & 30 & 0.5 & 3000 \\
Cluster - A & 1000 & 1 & 0.01 & 1100 \\
Cluster - B & 100 & 1 & 0.01 & 3700 \\
\hline
\hline
\hline 
\end{tabular}\end{center}
\caption{\label{table:impact} Characteristic number densities, velocities, time intervals and minimum 
impact parameters for several models of the local Solar neighborhood. We adopt the
parameters of 'Present Neighborhood - B' for our calculations.}
\end{table*}

\begin{figure}
\centering
\includegraphics[width=8cm,height=8cm]{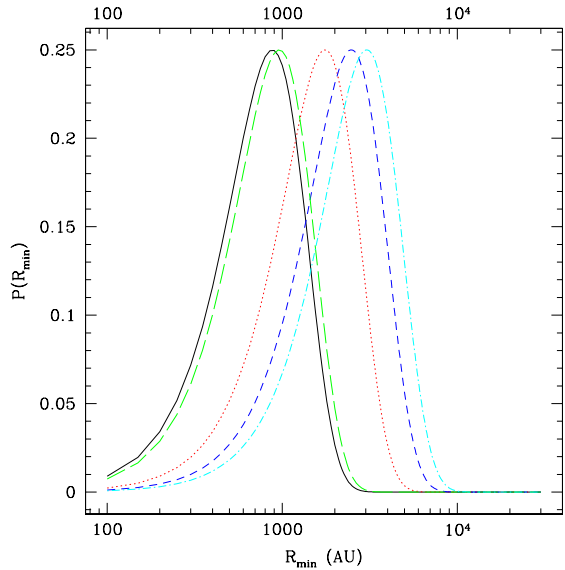}
\caption{\label{fig:impact} Plot of the probability distribution of minimum impact parameters for the models 
described in Table (\ref{table:impact}). The curves in a left-to-right manner correspond to Neighborhood-A 
(black, solid curve), Cluster-A (green, long dashed curve), Neighborhood-B (red, dotted curve), 
Neighborhood-C (blue, dashed curve) and Cluster-B (cyan, dot-dashed curve). We adopt the Neighborhood-B model 
for calculations in this paper.}
\end{figure}

Of course rare events ($P \ll 1$) sometimes occur and the above figures should be seen as estimates of the mean of 
the distribution from which the actual closest impact parameter has been randomly drawn. The probability that the 
closest encounter occurs with impact parameter $R$ is equivalent to the joint probability that one encounter occurred 
at $R$, $P(1|R)$, and no encounters occurred with $r < R$, $P(0|r<R)$,
\begin{align}\label{eq:prob}
P(R_{min}) =& P(0|r<R) \times P(1|R), \nonumber \\
 =&  \int_0^R p(r) dr e^{-\pi r^2 n v \Delta t} \times (\pi R^2 n v \Delta t) e^{-\pi R^2 n v \Delta t}, \nonumber \\
=&  e^{-\pi R^2 n v \Delta t}(1-e^{-\pi R^2 n v \Delta t}).
\end{align}
This probability has a maximum at $P = \ln{2} \sim 0.7$, instead of $P = 1$ in Eq. (\ref{eq:impact}).
The probability distribution of $R_{min}$ for the variety of environments are shown in Fig. \ref{fig:impact}.

\section{Analytic Estimates}
\label{sec:virial}

Before we present numerical results in \S \ref{sec:results} it is worthwhile to look at order-of-magnitude estimates 
of the effect. Using the impulse approximation it is straightforward to determine how the perturbation affects the 
test particles. The velocity and kinetic energy are directly changed by the impulse, but the potential energy remains 
the same since the distance from the Sun has not changed. If the system was self-gravitating and the traditional virial 
theorem applicable, the system would relax to a new equilibrium distribution and the moments of the gravitational potential 
tensor would directly reflect the changed velocity dispersion tensor \citep{Binney87}. Relating the velocity or energy 
kick to a quadrupole moment requires a generalization of the commonly used virial theorem as we show below.

\subsection{Virial Equations}

The ensemble of test particles in the Oort Cloud satisfies the collisionless Boltzmann equation
\begin{equation}\label{eq:cb}
\frac{\partial f}{\partial t} + {\bm v} \cdot \nabla f - \nabla \Phi \cdot \nabla_v f = 0,
\end{equation}
where the distribution function depends on position, velocity and time, although we will suppress this functional dependence 
for brevity. We will assume that the Oort Cloud is not self-gravitating, so the potential through which the test particles 
are moving is solely determined by the Sun's gravity
\begin{equation}
\Phi({\bm x}) = -\frac{G M_{\odot}}{r}.
\end{equation}
Assuming the distribution is time-independent, which is equivalent to assuming that Poisson fluctuations in the final moments are 
negligible (see \S \ref{sec:poisson}), we find the constraint equation
\begin{equation}\label{eq:boltzmann}
G M_{\odot} \frac{{\bm x} \cdot \nabla_v f}{r^3} = {\bm v} \cdot \nabla f.
\end{equation}
In simple words this equation states that the number of test particles advected from a phase space volume element equals the 
number of test particles accelerated into this volume element by the Sun's gravity. If we were deriving the traditional virial 
theorem we would take velocity and position moments of this equation or the time-dependent version of it, Eq. (\ref{eq:cb}). 
However we are interested in relating the velocity dispersion tensor to the intensity quadrupole moments, not the potential
energy tensor.

The quadrupole moment can be expressed as an integral over the Oort Cloud volume
\begin{align}\label{eq:quad_virial}
Q_{ij}(\nu) B_{\nu}(T_{\rm CMB}) =& \int d^3{\bm r} d^3{\bm v} f \frac{f_A}{r^4}(x_i x_j - \frac{1}{3}r^2 \delta_{ij}) \nonumber \\
&\times [B_{\nu}(T_{\rm Oort}) - B_{\nu}(T_{\rm CMB})].
\end{align}
The optical depth and the distribution function are defined such that
\begin{equation}
\int d^3{\bm r} d^3{\bm v} f({\bm x},{\bm v}) = 1.
\end{equation}
The blackbody frequency spectrum is in general a complex function of $T_{\rm Oort}$, although at low frequencies 
it scales as $B_{\nu}(T_{\rm Oort}) \propto T_{\rm Oort}$. In order to derive analytic results we will express 
the blackbody spectrum's dependence on distance as a power law,
\begin{equation}
B_{\nu}(T_{\rm Oort}) = \frac{A_{\nu}}{r^{\beta}},
\end{equation}
where the exponent is defined as
\begin{equation}
\beta = \frac{1}{2} \frac{\partial \ln{B_{\nu}(T)}}{\partial \ln{T}}.
\end{equation}
This exponent is only well defined in Rayleigh-Jean's limit where it equals a constant, $\beta = 1/2$. At higher frequencies
its value grows rapidly and is strongly frequency dependent. Below we will use $\beta = 1/2$, but comment on how the higher 
frequency cases will affect our results.

We first multiply Eq. (\ref{eq:boltzmann}) by ${\bm v} {\bm x}
r^{-(1+\beta)}$. Now following the basic steps in deriving the virial
theorem we will integrate this equation over phase space and then integrate
by parts to eliminate the position and velocity derivatives acting on the
distribution function. This will allow us to relate the first part, the
$x_i x_j$ term, of the quadrupole moment to the velocity dispersion tensor. The second
part, the $-\delta_{ij}/3$ term, can be related to the velocity through the
energy equation of a Keplerian orbit
\begin{equation}
E = \frac{v^2}{2} - \frac{G M_{\odot}}{r} = -\frac{G M_{\odot}}{2a}.
\end{equation}
For simplicity, here and in the next subsection we assume zero eccentricity orbits with $v^2 = G M_{\odot}/a$.

The final equation relating the quadrupole moments to the velocity distribution of the test particles is
\begin{align}\label{eq:virial_eq}
&Q_{ij}(\nu) B_{\nu}(T_{\rm CMB}) = f_A \int d^3{\bm r} d^3{\bm v} f({\bm x},{\bm v}) \{ A_{\nu} \nonumber \\ 
&\times \left[\frac{v_i v_j}{G M_{\odot}r^{1+\beta}} - \frac{(2+\beta) (x_i v_j + v_i x_j)({\bm x}\cdot{\bm v})}{2 G M_{\odot}r^{3+\beta}}
- \frac{\delta_{ij}}{3} \left( \frac{v^2}{G M_{\odot}}\right)^{2+\beta} \right] \nonumber \\
&- B_{\nu}(T_{\rm CMB})
\left[\frac{v_i v_j}{G M_{\odot}r} - \frac{(x_i v_j + v_i x_j) ({\bm x}\cdot{\bm v})}{G M_{\odot}r^{3}}
- \frac{\delta_{ij}}{3} \left( \frac{v^2}{G M_{\odot}}\right)^{2}\right] \}.
\end{align}

\subsection{Perturbations}

In this subsection we will make several approximations that will allow us to calculate analytic expressions for 
the quadrupole moments. We will adopt the impulse approximation as well as the tidal approximation for the
relative acceleration of the test particles with respect to the Sun
\begin{equation}
\Delta {\bm a}_{rel} = \frac{G M_p}{b^3}[{\bm r} - 3 \hat{\bm b} ({\bm r} \cdot \hat{\bm b})].
\end{equation}
In this expression for the relative acceleration, $M_p$ is the perturber mass, ${\bm r}$ is the radial vector between
the Sun and the Oort Cloud object, and ${\bm b}$ is the impact parameter vector between the Sun and the perturber.
We are able to use the tidal approximation in this section because we will assume a thin shell distribution 
of semi-major axes for all the test-particles. In our numerical calculations we assume a power-law distribution 
of semi-major axes between $a = 10^3 \au$ and $10^4 \au$. With this latter distribution our analytic model 
assumption of the tidal acceleration is invalid because for the appropriate values of $R_{\rm min}$ the perturber 
passes through the Oort Cloud.

With these assumptions for the analytic model, the velocity kick is,
\begin{equation}
\Delta {\bm v} = \frac{2 G M_p}{B^2 v_{rel}} ({\bm I} - \hat{\bm v}^p \hat{\bm v}^p) 
[{\bm r} - 3 \hat{\bm b} ({\bm r} \cdot \hat{\bm b})],
\end{equation}
where we project out the component of the force parallel to the perturber's
trajectory.  For later convenience we will define
\begin{equation}
{\bm \Psi} = \hat{\bm n} - 3 \hat{\bm b} (\hat{\bm n} \cdot \hat{\bm b}) - \hat{\bm v}^p (\hat{\bm n} \cdot \hat{\bm v}^p),
\end{equation}
such that
\begin{equation}
\Delta {\bm v} = \frac{2 G M_p r}{b^2 v_{rel}} {\bm \Psi}. 
\end{equation}
Here we have assumed
\begin{equation}\label{eq:phase}
\hat{\bm v}_{rel} \cdot \hat{\bm b} = 0,
\end{equation} 
which directly follows from our use of the impulse approximation. Working within the impulse approximation the direction 
of the impact parameter is almost entirely determined by the direction of the encounter velocity.
This constraint from the impulse approximation allows us to simplify the form of the above velocity kick. There is 
still phase information that is needed to completely specify the geometry, Eq. (\ref{eq:phase}) is simply one equation 
for two unknowns. For example, the direction of the encounter only determines a plane within which the direction of the 
encounter velocity must lie; the azimuthal angle of the encounter velocity must still be specified. Equivalently, the 
direction of the encounter velocity determines a great circle on the sky along which the encounter must have occured; 
the longitudinal position along that great circle must still be specified. The azimuthal angle of the relative velocity 
should have uniform distributions between $0$ and $\pi$ (recall the $\hat{\bm v} \rightarrow -\hat{\bm v}$ degeneracy 
mentioned above), whereas the longitudal angle that determines the direction of the impact parameter should be uniformly
distributed between $0$ and $2\pi$.

There is no change in the velocity dispersion, or the quadrupole moments,
at first order in $\Delta {\bm v}$.  This is a consequence of the fact that
we assumed the Oort Cloud to be statistically time-independent and
axisymmetric. At second order the velocity kick will produce a non-zero
quadrupole moment,
\begin{align}\label{eq:quad}
&Q_{ij}(\nu) B_{\nu}(T_{\rm CMB}) = \frac{4 G M^2_p \tau a^3}{b^4 v^2_{\rm vel} M_{\odot}} 
\int d^2\hat{\bm n} \{\frac{A_{\nu}}{a^{\beta}}[\Psi_i \Psi_j \nonumber \\
&- (1+\frac{\beta}{2})(\Psi_i n_j + \Psi_j n_i)(\hat{\bm n} \cdot \Psi) 
- \frac{(2+\beta)}{3}\delta_{ij} {\bm \Psi} \cdot {\bm \Psi}] \nonumber \\
&- B_{\nu}(T_{\rm CMB})[\Psi_i \Psi_j - (\Psi_i n_j + \Psi_j n_i)(\hat{\bm n} \cdot \Psi) - \frac{2}{3}\delta_{ij} 
{\bm \Psi} \cdot {\bm \Psi}]\},
\end{align}
where we have assumed that the Oort Cloud is made of a thin shell of zero eccentricity test particles all with 
semi-major axis $a$. Since we assume that all of the Oort Cloud objects lie at a single semi-major axis $a$
we can replace the absolute covering fraction $f_A$, defined in Eq. \ref{eq:opt}, with the optical depth $\tau$.

The quadrupole tensor can now be directly evaluated as
\begin{align}\label{eq:quad_direct}
Q_{ij}(\nu) = \frac{16 \pi G M^2_p \tau a^3}{15 b^4 v^2_{\rm vel}}  
\left[\frac{B_{\nu}(T_{\rm Oort})}{B_{\nu}(T_{\rm CMB})} A_{ij}(\beta) - A_{ij}(0)\right],
\end{align}
where the configuration tensor is defined as
\begin{equation}
A_{ij}(\beta) = [(3 - \beta)v_i v_j + 3(1 + 3 \beta) b_i b_j] + \frac{41 + 28\beta}{3}\delta_{ij}.
\end{equation}
Note that $v_i$ and  $b_i$ are components of $\hat{\bm b}$ and $\hat{\bm v}^p$, which have unit norm.

Taking some characteristic numbers, such as $M_p = 1 M_{\odot}$, $\beta=1/2$,\footnote{For simplicity we choose to evaluate the 
configuration tensor at $\beta =1/2$ for all frequencies despite the fact that $\beta$ and therefore $Q$ will grow at higher 
frequencies.} $v_{\rm rel} = 30 \kms$, $a = 10^3 \au$ and $b = 2 \times 10^3 \au$, we find
\begin{equation}
Q(\nu) = 3.9 \times 10^{-9} \left(\frac{\tau}{10^{-6}}\right) \left[ \frac{B_{\nu}(10 \K)}{B_{\nu}(3 \K)} - 1 \right].
\end{equation}
Ignoring the directional dependence described above, the amplitude of $Q$ varies from $1.4 \times 10^{-8}$ at low frequency 
to $2.3 \times 10^{-7}$ at $\nu = 300 \GHz$ and $8.9 \times 10^{-6}$ at $\nu = 600 \GHz$. Our choice of $\tau = 10^{-6}$
is close to the value of our adopted model in Table \ref{qaz}. Since we have assumed that initial distributions of the 
test particle sizes and orbital elements are independent, we can simply scale the value of $Q(\nu)$ to the appropriate
value of $\tau$. 

\section{Numerical Results}
\label{sec:results}

Next we present our numerical results. We start by randomly choosing the initial orbital elements of $10^6$ Oort Cloud 
objects according to the distributions in Eqs. (\ref{eq:ic}). Then for a given trajectory and mass of the external perturber, 
we use Eqs. (\ref{eq:pert}) to map the initial orbital elements to the final orbital elements. The resulting quadrupole moments, 
calculated using Eq. (\ref{eq:real_quad}), are determined for $10^3$ distinct initial and final times in order to estimate the 
effect of statistical fluctuations on the outcome and therefore determine if the perturbation produced a statistically significant 
change in the structure of the Oort Cloud. 

\begin{figure}
\centering
\includegraphics[width=8cm,height=8cm]{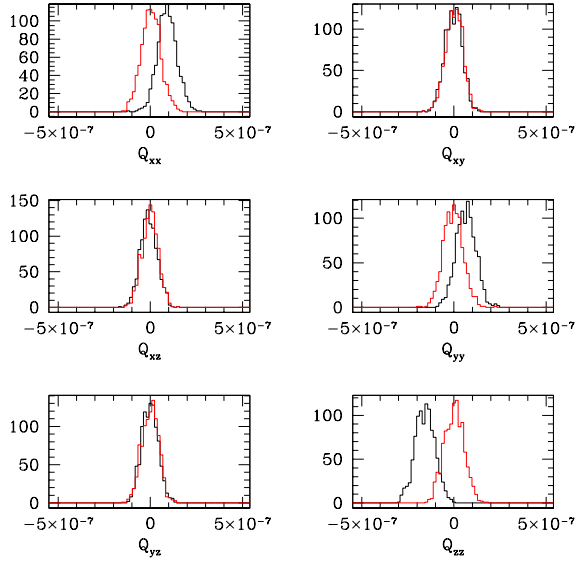}
\caption{\label{fig:pert_30} Histogram of the induced CMB quadrupoles at $\nu = 30 \GHz$ 
before (red curve) and after (black curve) the perturbation. The quadrupole moments are 
in units of mean CMB intensity.}
\end{figure}

\begin{figure}
\centering
\includegraphics[width=8cm,height=8cm]{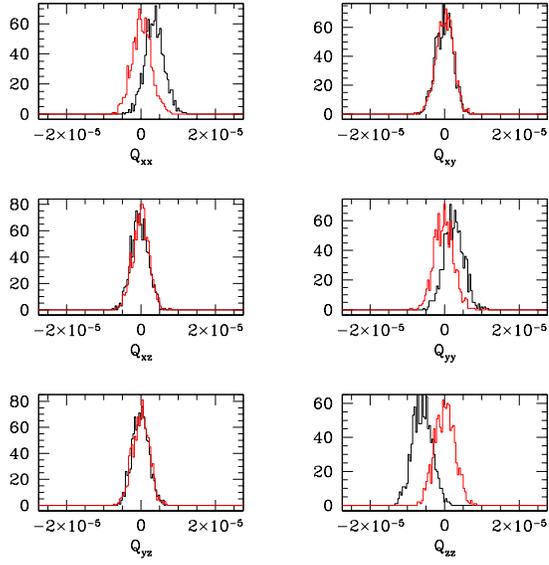}
\caption{\label{fig:pert_300} Same as Fig. (\ref{fig:pert_30}), but at $\nu = 353 \GHz$.}
\end{figure}

\begin{figure}
\centering
\includegraphics[width=8cm,height=8cm]{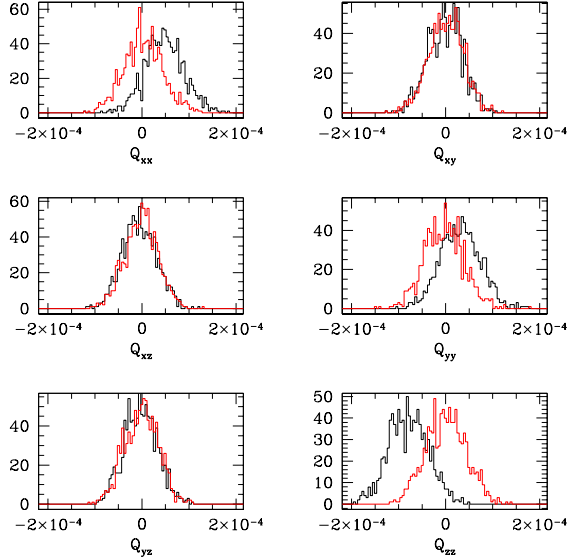}
\caption{\label{fig:pert_600} Same as Fig. (\ref{fig:pert_30}), but at $\nu = 545 \GHz$.}
\end{figure}

These quadrupole distributions are shown in Fig. \ref{fig:pert_30} for $\nu = 30 \GHz$, Fig. \ref{fig:pert_300} for $\nu = 353 \GHz$ 
and Fig. \ref{fig:pert_600} for $\nu = 545 \GHz$. We choose these frequencies to correspond to the frequency bins of the Planck satellite. 
The statistical 
fluctuations are greater at high frequency because the induced signal scales as a stronger power of radial distance, as seen in 
Eq. (\ref{eq:scale}). For this particular example the direction of impact parameter was $\hat{\bm B} = (0.909,0,-0.416)$ and the 
trajectory of the perturber was $\hat{\bm v}_p = (0,1,0)$. The perturber's mass was $M_p = 1 M_{\odot}$, its relative velocity 
$v_p = 30 \kms$ and impact parameter $b = 2100 \au$.

From the distributions of the quadrupoles in Figs. (\ref{fig:pert_30}-\ref{fig:pert_600}), 
we can determine that the non-zero value of the induced quadrupoles is statistically
significant. Although our numerical calculations suffer from Poisson fluctuations, the 
signal from the real Oort Cloud will not exhibit the relatively large changes over an 
orbital period if our model for the size distribution of Oort Cloud objects and estimate 
of the total mass are reasonable. The quadrupole moments shown in Figs. (\ref{fig:pert_30} 
- \ref{fig:pert_600}) can be converted to a value of $C_2$, the Fourier space quadrupole 
power spectrum, using the formulae derived in Appendix \ref{sec:quad}. The black points in
Fig. \ref{fig:power} are the quadrupole power spectrum, $C_{\ell=2}$, as a function of 
frequency along with the expected Planck instrumental noise errorbars\footnote{The 
instrumental error bars appear anomalously low because we are accustomed to seeing the 
cosmic variance errors bar dominate at low $\ell$.}. The expected statistical fluctuations
due to Poisson fluctuations in the number of observed Oort Cloud objects are much smaller 
for the models considered in this paper (see Table \ref{table:poisson}). Of course, the observability
of the signal will primarily be affected by all of the astrophysical sources that emit non-blackbody
radiation, such as the standard Galactic foregrounds, thermal Sunyaev-Zel'dovich emission, etc.
The signal originating in our Solar System must be observed against this background. The full 
calculation of the power spectrum as function of frequency of these various backgrounds is beyond 
the scope of this paper. 

Also shown in Fig. \ref{fig:power} is the quadrupole power spectrum of 
the Scattered Disc (blue points) and the theoretical power spectrum predicted by the WMAP 
cosmology model (red points). The theoretical power spectrum is frequency independent when 
expressed in terms of temperature fluctuations and has a value of $C_2 = 1.76 \times 10^{-10}$ 
\citep{Hinshaw06}. Since we plot the intensity fluctuations, and not the temperature fluctuations, 
the theoretical power spectrum displays a slight frequency dependence.

\begin{figure}
\centering
\includegraphics[width=8cm,height=8cm]{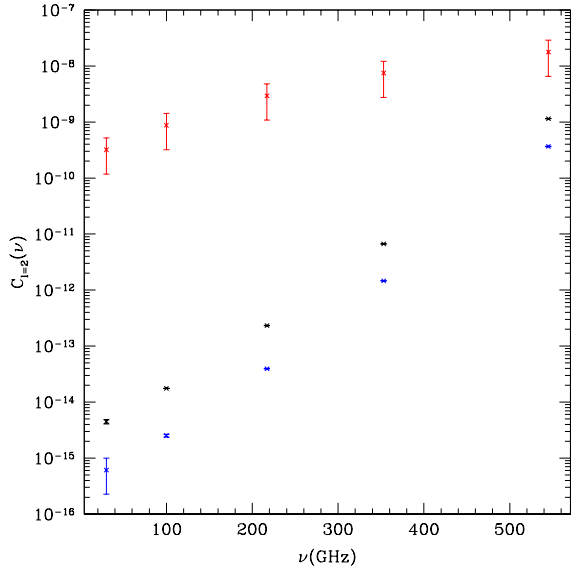}
\caption{\label{fig:power} The quadrupole power, $C_{\ell=2}$, at several different frequencies.
The power spectrum is dimensionless since the Oort Cloud quadrupole moments are expressed in terms
of the mean CMB intensity. The black points are the Oort Cloud power spectrum with error bars 
appropriate for Planck instrument noise. The blue points are the power spectrum produced by our 
model of the Scattered Disc. The red points are the theoretical power spectrum calculated with 
the best fit WMAP cosmology, corresponding to $C_2 = 1.76 \times 10^{-10}$ in terms of temperature 
fluctuations, with cosmic variance error bars. Since we chose to represent the fluctuations in 
terms of intensity and not temperature, the primordial temperature anisotropies do not produce 
a frequency independent power spectrum.}
\end{figure}

Our assumption that the distributions of test particle mass and orbital elements are independent allows us to vertically scale the 
curves in Fig. \ref{fig:power} as we change $f_A$. The predictions for the power spectrum scale as $f^2_A$. Although we have 
separately plotted the quadrupole power spectra for the Oort Cloud and the Scattered Disc, in reality we will observe the combination 
of the two. We chose to separate the two contributions in Fig. \ref{fig:power} because the Scattered Disc is intrinsically aspherical 
and the Oort Cloud's asphericity is induced by a perturbation. Even if the Oort Cloud was perfectly spherical, the scattered disc, 
which is believed to smoothly compress from a spherical shape at the inner edge of the Oort Cloud into the Ecliptic Plane at the 
outer edge of the Kuiper Belt, will produce a minimum quadrupole moment. 

While we calculate the corresponding quadrupole power spectrum values for comparision with CMB values, we should emphasize 
that the individual quadrupole moments of the induced spectral distortions contain phase information. It is assumed that 
the CMB temperature anisotropies constitute a Gaussian random field and the observed $a_{\ell m}$'s are interpreted as being 
a random realization drawn from a Gaussian probability distribution function with variance $C_{\ell}$. In which case, the only 
meaningful information in a set of CMB data are the measured values of the $C_{\ell}$'s. For the anisotropic spectral distortions 
induced by the Oort Cloud the particular values of the individual $a_{2 m}$'s, or equivalently the $Q_{i j}$ (see Appendix \ref{sec:quad} 
for relations that link these quantities), contain important information about the structure of the Oort Cloud. This is an important 
distinction to remember and one that we will explore in more detail in \S \ref{sec:bayes}.

The instrumental noise error bars appear anomalously small in Fig. \ref{fig:power}. Typically on large scales cosmic variance 
dominates instrumental noise; for Planck the power spectrum should be cosmic variance limited until $\ell \approx 1500$. 
The error bars on the theoretical primordial quadrupole power spectrum in Fig. \ref{fig:power} are due to cosmic variance. 
Although the Oort Cloud and Scattered Disc power spectra are never distinguishable from the theoretical power spectrum at a single
frequency, we can use the fact the primoridal CMB temperature anisotropies are frequency independent to utilize the low frequency 
data (where the Oort Cloud signal is effectively zero) to eliminate the cosmic variance uncertainity in the primordial temperature
anisotropies at high frequency. Namely the low frequency temperature anisotropies can be used to exactly predict the high frequency 
temperature anisotropies\footnote{This statement depends on the level of foreground contamination in the low frequency bands}. 
The fact that cosmic variance limits our ability to measure the true power spectrum is irrelevant because we only care about 
the particular realization of the temperature anisotropies in our Universe. This should allow us to use the temperature anisotropies 
in a single frequency bin to remove them from the other frequencies and detect the Oort Cloud signal whenever it is larger 
than instrumental noise.

\section{Bayesian Analysis}
\label{sec:bayes}

The observed CMB can be used to constrain the parameters of the stellar encounter if a quadrupole, or any higher order moment, 
spectral distortion were to be detected. The {\it posterior} likelihood of the Oort Cloud and external perturbation parameters, 
collectively designated as $T$, given the observed data, designated as $D$, is
\begin{equation}
\mathcal{L}(T|D) \propto \mathcal{L}(D|T) \mathcal{L}(T),
\end{equation}
where we have used Bayes' Theorem to invert the conditional probabilities. The external perturbation parameters depend on the 
mass of the perturber, $M_p$; the impact parameter of the encounter, ${\bf b}$; the velocity of the perturber with respect to 
the Solar System, ${\bf v}_{rel}$; and the absolute covering fraction of Oort Cloud objects, $f_A$. There exist degeneracies 
between some of these parameters. For example, the projection matrix, defined in Eq.(\ref{eq:proj}), is invariant under the 
change $\hat{v} \rightarrow -\hat{v}$ since the force parallel to the trajectory cancels in the impulse approximation, and 
the total impulse only depends on the combination $M_p/b^2 v_{rel}$. Plots of the parameter constraints will directly display 
these degeneracies as the data is only sensitive to particular combinations, although the {\it a priori} probabilities will 
constrain these parameters to some extent.

The likelihood function of the data given the stellar encounter parameter can be expressed as a functional integration over the 
true CMB signal (S), which will be a combination of primordial temperature anisotropies and Oort Cloud emission,
\begin{equation}
\mathcal{L}(D|T) = \int dS \mathcal{L}(D|S)\mathcal{L}(S|T),
\end{equation}
where the likelihood of the data given the true CMB signal is simply set by instrument noise \citep{Knox95,Dodelson03}
\begin{equation}
\mathcal{L}(D|S) = \frac{1}{\sqrt{2\pi \det{N}}}e^{-{\bm (d-a)}^{\dagger}{\mathbf N}^{-1}{\bm (d-a)}/2},  
\end{equation}
where ${\mathbf N}$ is the noise covariance matrix, $d$ is the observed signal and $a$ is the true underlying CMB signal.
The observed signal is not uniquely determined by the underlying theory which
only specifies the probability distribution function of which the signal is a random realization. The likelihood function
of the CMB temperature anisotropies is
\begin{equation}
\mathcal{L}(S|T) = \frac{1}{\sqrt{2\pi \det{C}}}e^{-{\bm a}^{\dagger}{\mathbf C}^{-1}{\bm a}/2},  
\end{equation}
where ${\mathbf C}$ is the signal covariance matrix. The signal covariance matrix is defined as the expectation value 
of ${\mathbf C} = \langle {\bm a} {\bm a}^{\dagger} \rangle$, with a similar definition for the noise covariance matrix 
${\mathbf N}$. The likelihood of the Oort Cloud signal given the theory is set a probability distribution is determined 
by Poisson fluctuations as discussed above in \S \ref{sec:poisson}. 

The {\it a priori} probability of Oort Cloud and stellar perturbation parameters should not be modeled as the product of the 
individual parameter distributions. Observations of the local Solar neighborhood have found that different stellar spectral 
types have different number densities and different velocity ellipsoids \citep{Binney98,King90}. We should be mindful that
the Sun's local environment might have been quite different in the past as the Sun moves through different Galactic environments.
The {\it a priori} distribution can be written in terms of conditional probabilities as
\begin{equation}
\mathcal{L}(D) = p(b|v_{rel},M)~p(\hat{b}|\hat{v})~p({\bf v}_{rel}|M)~p(M)~p(f_A),
\end{equation}
where p(M) is the distribution of stellar masses, averaged according to stellar evolutionary theory as
\begin{equation}
p(M) = \Psi(M) \frac{t(M)}{\Delta t},
\end{equation}
where $\Psi(M)$ is the appropriate initial mass function, $t(M)$ is the lifetime of stars of mass $M$ and
$\Delta t$ is the lifetime of the Solar System. The distribution of velocities is assumed to follow a 
Schwartzschild distribution \citep{Binney87}
\begin{equation}
p({\bf v}_{rel}|M) = \frac{1}{(2\pi)^{3/2} \det{{\mathbf \Sigma}}}e^{- {\mathbf v} {\mathbf \Sigma} {\mathbf v}/2}
\end{equation}
with the velocity ellipsoid (${\mathbf \Sigma}$) depending on the stellar spectral type \citep{King90,Binney98}. As discussed 
in \S \ref{sec:virial}, the direction of the impact parameter is the determined by the direction of the encounter velocity up to 
a phase angle. The magnitude of the impact parameter depends on the number density of stars of a given mass and their relative 
velocity with respect to the Sun\footnote{Gravitational focusing will induce an additional correlation between the distribution 
of $b$ and $v_{rel}$, but we ignore it here since gravitational focusing is not consistent with our use of the impulse approximation.}.
Given these parameters, the distribution of impact parameters can be calculated as outlined in \S \ref{sec:impact}. The {\it a priori}
probability of $f_A$ is completely unknown, except that an upper limit can be set using constraint on the mean CMB spectrum distortion
\citep{Babich06}. We can consider the CMB power spectrum, needed to specify the likelihood function of the extragalactic temperature
anisotropies, as known.

If a quadrupole spectral distortion in the CMB is definitively measured then this formalism can be exploited to constrain the closest 
stellar encounter. There are similiar hopes to constrain such a stellar perturbation by using the extreme orbits of Sedna and analogs 
that should be discovered in the future. It remains to be seen if this goal can actually be accomplished as the associated CMB spectral 
distortion moments suffer from large statistical fluctuations and the initial conditions of the Sedna-like objects are unknown.

\section{Conclusions}
\label{sec:concl}

We analyzed the effect of a close stellar encounter on the shape of the Oort Cloud and the subsequent anisotropic CMB spectral distortions 
that would result. Much about the Oort Cloud is unknown and our calculations in this paper are subject to these uncertainities. We have 
strived to adopt conservative estimates for our mass model of the Oort Cloud and have highlighted the resulting uncertainities. The most 
likely minimum impact parameter over the past $1 \Gyr$ for the present-day local Solar neighborhood is $b = 2100 \au$. Assuming the perturber 
is a Solar mass star traveling with relative velocity $30 \kms$ with respect to the Sun, the perturbation will produce frequency-dependent 
quadrupole moments in the observed CMB of $C_2 = 4.5 (3.5) \times 10^{-15}, 6.7 (1.1) \times 10^{-12}, 1.1 (0.11) \times 10^{-9}$ at 
$\nu = 30, 353, 545 \GHz$, respectively in the intensity and (temperature) fluctuations. The asphericity in the Oort Cloud that leads to our 
signal will produce spectral distortions in the CMB on all angular scales. Our calculation can be extended to the octopole ($\ell = 3$), or 
any higher moment, without much difficulty. Moreover the signal will be highly non-Gaussian and will induce correlations between the power 
spectra on different scales. These features may help in the detection of the signal. Using data from scales smaller than quadrupole will also 
help in disentangling the signal from the Galactic foreground emission.

The Planck Satellite\footnote{www.rssd.esa.int/Planck/} will have nine bands of frequency coverage between $\nu = 30 \GHz$ and $857 \GHz$, 
with the $\nu = 353 \GHz$ band the highest with reasonable estimated detector noise.  Additionally, the recently launched far-infrared 
satellite Akari, originally Astro-F, has frequency coverage down to $2 \THz$ \citep{Pearson01} and might be able to search for the signal 
from the closest and hottest Oort Cloud objects.

Finally, we described a Bayesian method to use observed anisotropic spectral distortions to constrain the trajectory of the closest stellar
perturber. The Bayesian method requires that the direction of the quadrupole, or equivalently the phase information in the quadrupole
moments, does not drastically change over timescales comparable to the period of the Oort Cloud objects. The associated fluctuations are due 
to Poisson fluctuations in the number of Oort Cloud objects at any position in the outer Solar System. We quantified these fluctuations and 
determined that they are important when interpreting our calculations due to the relatively small number of test particles ($10^9$) we used. 
However, we estimated that the real Oort Cloud has approximately $10^{28}$ objects and the interpretation of the real quadrupole moments would 
not be affected by these statistical fluctuations. Therefore the direction of the quadrupole of CMB spectral distortions, if ever observed, 
could be used to constrain the properties of the closest stellar encounter with the Solar System.

{\bf Acknowledgements}

We would like to thank M. Brown, P. Goldreich, M. Holman,
Z. Lienhardt and P. Weissman for helpful conversations and L. Dones for
sharing an unpublished manuscript.  DB thanks the hospitality of the
Harvard Institute for Theory and Computation where some of this work was
completed and acknowledges financial support from the Betty and Gordon
Moore Foundation.

\appendix

\section{Angular Moment Definitions}
\label{sec:quad}

In this appendix we review the relationships between the various
descriptions of the angular moments of a distribution on a sphere, in
particular the quadrupole moments which are discussed in this paper.  It is
advantageous to describe the CMB temperature anisotropies in a basis of
spherical harmonics because of their orthonormality properties on the
sphere. A function defined over the sphere, $A(\hat{\bm n})$, can be
expanded in a spherical harmonic basis as
\begin{equation}
a_{\ell m} \equiv \int d^2\hat{\bm n} Y^{*}_{\ell m}(\hat{\bm n}) A(\hat{\bm n}).
\end{equation}
Since the CMB temperature anisotropies are assumed to be described by a Gaussian random field the 
particular values of the $a_{\ell m}$ are not meaningful, only the statistical properties of the 
expansion coefficients carry any information about the underlying cosmology. Therefore, it is customary 
to calculate the ensemble average of the $a_{\ell m}$ covariance matrix
\begin{equation}
\langle a_{\ell_1 m_1} a_{\ell_2 m_2} \rangle \equiv \delta_{\ell_1,\ell_2} \delta_{m_1, -m_2} (-1)^{m_1} C_{\ell_1},
\end{equation}
which can be related to an isotropic\footnote{Invariant under rotations of our coordinate system. For example, it is 
invariant with respect to different choices of our reference direction.} quantity, the power spectrum, $C_{\ell}$.
This identity can be used to motivate an observational definition of $C_{\ell}$
\begin{equation}\label{eq:sum}
 C_{\ell} = \frac{1}{2\ell + 1} \sum_{m=-\ell}^{\ell} (-1)^m a_{\ell m} a_{\ell -m}. 
\end{equation}

We will now relate these quantities to the Cartesian multipole moment expansions utilized in our 
paper and defined in Eq. (\ref{eq:real_quad}). Expressing the Cartesian directional vectors
in spherical coordinates,
\begin{align}
\hat{n}_x + i \hat{n}_y &= \sin{\theta} e^{i\phi}, \\
\hat{n}_z &= \cos{\theta}.
\end{align}
Then the five $\ell = 2$ spherical harmonics can be expressed as
\begin{align}
Y_{2 2}(\hat{\bm n}) &= \sqrt{\frac{3}{8}} \sqrt{\frac{5}{4\pi}} (\hat{n}_x + i \hat{n}_y)^2, \\
Y_{2 1}(\hat{\bm n}) &= - \sqrt{\frac{3}{2}} \sqrt{\frac{5}{4\pi}} \hat{n}_z (\hat{n}_x + i \hat{n}_y), \\
Y_{2 0}(\hat{\bm n}) &= \frac32 \sqrt{\frac{5}{4\pi}}(\hat{n}_z^2 - \frac13), \\
Y_{2 -1}(\hat{\bm n}) &= \sqrt{\frac{3}{2}} \sqrt{\frac{5}{4\pi}} \hat{n}_z (\hat{n}_x - i \hat{n}_y), \\
Y_{2 -2}(\hat{\bm n}) &= \sqrt{\frac{3}{8}} \sqrt{\frac{5}{4\pi}} (\hat{n}_x - i \hat{n}_y)^2.
\end{align}
From these definitions relationships between the $a_{2 m}$ and $Q_{i j}$ can be derived.
\begin{align}
a_{2 2} &= \sqrt{\frac{3}{8}} \sqrt{\frac{5}{4\pi}} (Q_{xx} - Q_{yy} - 2iQ_{xy}), \\
a_{2 1} &= -\sqrt{\frac{3}{2}} \sqrt{\frac{5}{4\pi}} (Q_{xz} - iQ_{yz}), \\
a_{2 0} &= \frac{3}{2} \sqrt{\frac{5}{4\pi}} Q_{zz}, \\
a_{2 -1} &= \sqrt{\frac{3}{2}} \sqrt{\frac{5}{4\pi}} (Q_{xz} + iQ_{yz}), \\
a_{2 -2} &= \sqrt{\frac{3}{8}} \sqrt{\frac{5}{4\pi}} (Q_{xx} - Q_{yy} + 2iQ_{xy}).
\end{align}
The power spectrum can be related to the Cartesian quadrupole moments through Eq. (\ref{eq:sum})
\begin{equation}\label{eq:quad_def}
 C_{2} = \frac{3}{4 \pi} (Q^2_{xx} + Q^2_{zz} + Q^2_{xy}+ Q^2_{xz} + Q^2_{yz} + Q_{xx}Q_{zz}).
\end{equation}
The expectation value of this expression should contain both the mean and variance of the $Q_{ij}$ 
as terms. However, consistent with our calculation that Poisson fluctuations are not important we 
will evaluate Eq. (\ref{eq:quad_def}) with only the means of the $Q_{ij}$.

\section{Instrument Noise}
\label{sec:noise}

Finally, we describe the instrument noise values that were adopted in this
paper. In temperature units the instrument noise for the observation of
$C_{\ell}$ is \citep{Knox95}
\begin{equation}
C^N_{\ell} = \sqrt{\frac{2}{2\ell+1}}\frac{4\pi\sigma^2}{N_{pix}} e^{\ell^2\sigma_B^2},
\end{equation}
where the beam width is related to the FWHM as $\sqrt{8 \ln{2}} \sigma_B = $FWHM, $N_{pix}$ is the
number of observed pixels and $\sigma$ is the noise per pixel in temperature units.
The temperature fluctuation can be related to an intensity fluctuations as follows,
\begin{equation}
\Delta I_{\nu} = \frac{\partial B_{\nu}(T)}{\partial T}|_{T_{\rm CMB}}\Delta T_{\nu}.
\end{equation}
Table \ref{table:noise} lists the noise parameters used in this paper.

\begin{table}
\begin{center}\label{table:noise}
\begin{tabular}{ccc}
\hline
\hline 
$\nu$ & $\sigma/T_{\rm CMB} \times 10^{-6}$ & $FWHM$ (arcmin) \\
\hline 
$30 \GHz$ & $2.0$ & $33$ \\
$100 \GHz$ & $2.5$ & $9.5$ \\
$217 \GHz$ & $4.8$ & $5.0$ \\
$353 \GHz$ & $14.7$ & $5.0$ \\
$545 \GHz$ & $147$ & $5.0$ \\
\hline
\hline 
\end{tabular}\end{center}
\caption{Instrument noise parameters adopted for this paper. These values
were taken from the Planck Science webpage (www.rssd.esa.int/Planck/).}
\end{table}

\end{document}